\documentclass[acmsmall,screen]{acmart}

\AtBeginDocument{%
  }

\setcopyright{acmlicensed}
\acmJournal{PACMHCI}
\acmYear{2024} \acmVolume{8} \acmNumber{CSCW2} \acmArticle{482} \acmMonth{11}\acmDOI{10.1145/3687021}

\usepackage{multirow}
\usepackage{makecell}
\usepackage{caption}
\usepackage{subcaption}

\begin{document}

\title{Anonymization of Voices in Spaces for Civic Dialogue: Measuring Impact on Empathy, Trust, and Feeling Heard}

\author{Wonjune Kang}
\authornote{Both authors contributed equally to this research.}
\email{wjkang@mit.edu}
\orcid{0000-0002-5363-706X}
\affiliation{%
  \institution{Massachusetts Institute of Technology}
  \city{Cambridge}
  \state{MA}
  \country{USA}
}

\author{Margaret A. Hughes}
\authornotemark[1]
\email{mhughes4@mit.edu}
\orcid{0009-0009-9046-4098}
\affiliation{%
  \institution{Massachusetts Institute of Technology}
  \city{Cambridge}
  \state{MA}
  \country{USA}
}

\author{Deb Roy}
\authornote{Deb Roy is part-time unpaid CEO of Cortico.}
\email{dkroy@mit.edu}
\orcid{0000-0002-2780-4768}
\affiliation{%
  \institution{Massachusetts Institute of Technology}
  \city{Cambridge}
  \state{MA}
  \country{USA}
}


\begin{abstract}
Anonymity is a powerful component of many participatory media platforms that can afford people greater freedom of expression and protection from external coercion and interference.
However, it can be difficult to effectively implement on platforms that leverage spoken language due to distinct biomarkers present in the human voice.
In this work, we explore the use of voice anonymization methods within the context of a technology-enhanced civic dialogue network based in the United States, whose purpose is to increase feelings of agency and being heard within civic processes.
Specifically, we investigate the use of two different speech transformation and synthesis methods for anonymization: voice conversion (VC) and text-to-speech (TTS).
Through a series of two studies, we examine the impact that each method has on 1) the empathy and trust that listeners feel towards a person sharing a personal story, and 2) a speaker's own perception of being heard, finding that voice conversion is an especially suitable method for our purposes.
Our findings open up interesting potential research directions related to anonymous spoken discourse, as well as additional ways of engaging with voice-based civic technologies.
\end{abstract}



\received{January 2024}
\received[revised]{April 2024}
\received[accepted]{May 2024}

\maketitle

\section{Introduction}

In recent years, civic engagement has been greatly driven by the use of digital and Internet-based participatory media.
Such participatory media include platforms such as social network services, blogs, podcasts, digital storytelling, and virtual communities.
Although all of these media have distinct characteristics, according to Rheingold~\cite{rheingold2008using}, they share the following commonalities.
First, they enable all participants to broadcast as well as receive information in many different modalities, including text, images, audio, and video.
Second, their value and power derives from the active participation of many people, and the links between users form a public as well as a market.
Third, when amplified by information and communication networks, they enable broader, faster, and lower cost coordination of a wide range of activities.
The key power to this form of media lies in its participatory potential, as it affords people the opportunity to express their own voice in such a way that allows them to turn their self-expression into a form of public participation~\cite{agre1999find}.
In particular, the public voices of individuals, when aggregated and put into dialogue with the voices of others, make up a fundamental component of the ``public sphere,'' as defined by the political philosopher Jürgen Habermas~\cite{habermas1991structural}.
Given the power and freedom to influence policy, the public sphere can become an essential instrument of democratic self-governance~\cite{kellner2000habermas}.

In these settings, ``voice'' can be thought of as the unique style of personal expression that distinguishes one's communications from those of others~\cite{rheingold2008using}.
At the same time, a person's literal voice has certain qualities that make it particularly effective for the purposes of participatory media described above.
In addition to linguistic information, the human voice contains a significant amount of paralinguistic information such as emotion, emphasis, contrast, and focus~\cite{tannen1982spoken}.
It can also encode other elements of self-expression that may not be encoded by simple grammar or vocabulary, such as irony or sarcasm~\cite{hirst1998survey}.
These factors make speech a much richer medium for communication than other modalities such as written text, and have contributed to rapidly growing interest in voice-based platforms such as Discord, as well as short-form video platforms such as TikTok, Instagram Reels, and YouTube Shorts, where voice can play a significant role in sharing information~\cite{gao2022impacts}.

In many such participatory media platforms, providing \textit{anonymity} to the voice can be a critical component of the overall system.
Anonymity affords certain freedoms to participants in civic discourse by providing protection from coercion and interference~\cite{Asenbaum2018AnonymityDemocracy}.
Additionally, it can promote greater openness of discussion by allowing people to express and discuss potentially controversial opinions and arguments without risk to their image or career.
One way of providing anonymity within social discourse is by following the so-called ``Chatham House rule''~\cite{house2017chatham}, in which participants in a meeting are ``free to use the information received, but neither the identity nor the affiliation of the speaker(s), nor that of any other participant, may be revealed.''
In practice, this is often done by simply taking a transcript or an abbreviated textual summary of the spoken content, censoring out any personally identifiable information, and further processing it for downstream usage of the information.

However, in settings where people are speaking and sharing powerful stories or life experiences, directly hearing their voices can be much more impactful than reading text summaries or transcripts; listening to someone tell a narrative can convey its essence much more effectively than simply reading a written version or transcript of the same story~\cite{mchugh2012affective}.
At the same time, speech contains a significant amount of personal information about the speaker, and it is a distinct biomarker by which a speaker's identity may be determined~\cite{nautsch2019preserving}.
This poses the question: \textit{How might we preserve the nuance, emotion, and other rich information encoded in speech while simultaneously providing anonymity to a speaker?}

Motivated by the above, this work explores the use of \textit{voice anonymization} methods, where the objective is to preserve the qualities of an original speaker's voice that convey rich paralinguistic information while masking the speaker's identity.
In particular, we study the application of voice anonymization in the context of a \textit{civic dialogue network} based in the United States, which consists of collections of multi-party conversations in which participants engage in dialogue with one another and share stories and life experiences in relation to community or social issues.
A key component of the civic dialogue network is the release of particularly insightful or powerful stories to the broader public and committed leadership figures (see Section \ref{sec:civic_dialogue_network}).
However, there exist cases where speakers may fear retribution or the release of sensitive information, which motivates the usage of a voice anonymization system that can mask their identities.

We consider two different speech transformation/synthesis methods for voice anonymization: voice conversion (VC) and text-to-speech (TTS).
In particular, we utilize state-of-the-art neural network-based methods~\cite{kang23b_interspeech, kim2021conditional} that enable high-quality manipulation and synthesis of voices in a way that was not possible only a few years ago.
We perform two studies to measure the impact of voice anonymization within our civic dialogue network.
In the first study, we survey participants using a crowdsourcing website ($n = 1500$) to investigate how anonymization using VC and TTS affects the perceptions of listeners towards a storyteller, especially in terms of empathy and trust.
We also explore whether informing listeners that audio has been altered for the purposes of anonymization affects their responses.
In the second study, we investigate speakers' perceptions of their own stories that have been anonymized via VC or TTS by recruiting actual participants in the civic dialogue network ($n = 21$).
We perform anonymization on audio clips of stories they shared during real-life dialogues and survey how anonymizing speech affects the speakers' impressions of how their emotions and intended message are conveyed.
We also study the extent to which they feel their identity has successfully been masked, and whether anonymization changes the different parties in their communities that they would be comfortable sharing their story with.

While a large body of prior research has studied the qualities of synthetic voices in human-to-speech interface settings~\cite{clark2019state}, to the best of our knowledge, there has been no work studying the impact of using these technologies to alter how real peoples' voices and stories are conveyed.
Our work contributes to the literature by exploring not just a novel use of speech transformation/synthesis technologies, but also by studying how they might be used to foster and enable greater civic discourse.
While we believe voice and anonymization in democratic participation to be relevant for exploration in democracies globally, due to the civic dialogue network we consider being positioned in the United States, our work primarily focuses on this specific context.
\section{Context and Related Work}
\label{related_work}

\subsection{The Role of Vocal Qualities in Spoken Language}
\label{related_work_1}

In spoken language, much of the meaning is determined by context---that is, the objects or entities which surround a focal communicative event~\cite{tannen1982spoken}.
This means that the truth or validity of a proposition is determined by commonsense reference to experience.
Consequently, spoken language tends to convey subjective information, and an important aspect of it is the establishment of a relationship between the speaker and the audience~\cite{tannen1982spoken}.
This contrasts with written language, in which there is a greater emphasis on logical and coherent argument and most of the meaning is provided directly by the text itself.
These properties of speech make it an ideal medium for people to share content such as stories and life experiences.
Speech allows a listener to pick up on paralinguistic cues that convey the valence of emotional experiences~\cite{mcaleer2014you, scherer2001emotion}, and it has been shown to communicate human-like mental capacities related to thinking and feeling~\cite{schroeder2015sound}.

Key to the listening experience is the role of \textit{prosody}; that is, how a given speaker uses patterns of intonation, stress, and rhythm to provide variations in their speech.
Prosody plays an important role when we listen to a voice, as it is a major determinant of expressiveness, emotion, and naturalness, which are some of the most important traits in human communication~\cite{giles1973communicative, rodero2011intonation}.
It also makes up a significant component in human interaction~\cite{pittam1994voice}; prosody features have been shown to help segment discourse and provide an acoustic signal for listeners to better understand speech~\cite{sanderman1997prosodic, rodero2015principle}.
Thus, by contributing to the expressiveness of a speaker's message, prosody can draw attention to and aid with understanding.

In the past, prosody representation in artificially generated voices was poor, as traditional speech synthesis methods were not able to match real human voices in terms of basic quality and naturalness.
However, newer speech synthesis methods based on deep neural networks are now capable of generating speech that rivals that of humans in terms of naturalness and prosodic variation~\cite{kong2020hifi, kim2021conditional, casanova2022yourtts, kharitonov2023speak}.
This has led to the development of many more practically usable speech interfaces, such as personal voice assistants like Amazon's Alexa, Apple's Siri, or Google Assistant.

The rapidly increasing popularity and prevalence of such technologies have motivated the Human-Computer Interaction (HCI) and Computer-Supported Cooperative Work (CSCW) communities to extensively study speech interfaces~\cite{clark2019state}, especially the the role of prosody and vocal qualities on human perception of both real and synthetic voices.
For example, previous research has studied the role of expressivity on the perception of storytelling speech~\cite{montano2016role} or how people utilize voice quality and prosody in the identification of emotions and attitudes~\cite{grichkovtsova2012role}.
On the speech synthesis side, prior work has evaluated different types of TTS voices for long-form content~\cite{cambre2020choice} or compared TTS and human voices in a narrative advertising setting, measuring how modifying prosody affected listeners in terms of effectiveness, attention, concentration, and recall~\cite{rodero2017effectiveness}.
Other work has studied social implications and research challenges in designing voices for smart devices, treating voice assistants as social agents within a human-robot interaction framework~\cite{cambre2019one}.

Most of the above work addressed the qualities of synthetic and/or human voices in the context of human-to-human or human-to-speech interfaces.
However, to the best of our knowledge, there has been no previous work studying the the human-perceived impact of using speech transformation or synthesis technologies to change how real peoples' voices are conveyed to an audience.
Prior work on voice anonymization~\cite{pobar2014online, qian2017voicemask, fang2019speaker, han2020voice, patino2021speaker} has primarily focused on developing improved technical systems, with evaluation limited to ``hard'' metrics such as success at fooling speaker identification systems or mean opinion score (MOS) listening tests for quality and intelligibility~\cite{tomashenko2022voiceprivacy}.
In this work, we consider a novel setting and evaluation method for anonymization, looking more at its perceptual impact on speakers and listeners using more human-driven metrics such as empathy, trust, and the conveyance of emotions and meaning.

\subsection{Voice, Participation, and Anonymity in Democracy} \label{related_work_2}

``Voice'' is often used as a metaphor for participation within civics and democracies. Extensive political literature emphasizes the importance of participation in democracy, and explores various means through which civic actors may participate~\cite{arnstein1969ladder, fung2006varieties}. In particular, we look to deliberation and discourse within democracies as a key form of political participation. Habermas outlines a foundational model of discourse within democracies through the ``public sphere''~\cite{habermas1991structural}. In the public sphere, a public, or evolving group of people, gathers to discuss and debate through conversation issues of concern to them to form an opinion. These gatherings are social, can be broken into smaller groups to discuss various issues, and are intended to be rational with reasoned arguments. The public sphere described here requires that all people within a society have access to participation, and that they should be independent from any governing body or coercion. Furthermore, the public sphere should have the ability to hold the governing state to account.

However, theorists identify that this is not the case in any democracy globally; rather, we see an emergence of ``subaltern counter publics,'' or other versions of ``public spheres'' that are often smaller, created by those marginalized within a society, and centered around a shared identity in which that community can gather and discuss issues important to them free from fear of retribution~\cite{fraser2014rethinking, allen2023justice}. Anonymization within modern democracy is as old as the protected vote, and is seen as another way to protect minority voices and opinions from retaliation by an oppressive or dominating group~\cite{mill1966liberty, tocqueville2016democracy}. In the United States, anonymity is prevalent throughout democracy, be it through voting systems, campaign funding, or political protest~\cite{Asenbaum2018AnonymityDemocracy}. Especially for marginalized groups, or in cases where specific insights might directly challenge some powerful governing body and that body wishes to suppress that knowledge, anonymity can promise some protection. It can give both positive and negative freedoms in that it protects civic actors from coercion and interference, and as well as in that it creates space for identity development or adjustment~\cite{Asenbaum2018AnonymityDemocracy}. Yet, anonymity also comes with certain challenges in that it can allow citizens to share harmful or hateful speech without consequence~\cite{sunstein1995democracy}. Additionally, while in some cases anonymity can protect and uplift marginalized minority voices from retaliation, direct democracies with anonymous participation can also easily enable tyranny of the majority as well, with no social consequence for a majority continuously pursuing their own interests at the expense of the minority~\cite{tocqueville2016democracy}.

The science, technology, and society community has long explored the impact of anonymity and identity masking in digital spaces. Many observed that technology and the masking of identity in online spaces invites participants to construct a version of their identity that might be different from that in the offline world, having a liberating and positive impact, but also a potentially harmful one~\cite{baym2015personal, boyd2012politics, donath2002identity, turkle2011life}. The early days of online social networks saw a lively debate (which continues today) around identity, verification and trust, as well as the value and unique conditions that enable masked identity online. While some platforms were founded on the principle of known communities in which verified identity was at the root~\cite{boyd2012politics}, other platforms, such as Yik Yak and Reddit, were designed for various levels of anonymity, resulting in a wide range of social dynamics in these spaces. Reddit has been used as a respite for the lonely and can provide supportive spaces for mental health discussions, while it can also create spaces that support socially taboo behavior~\cite{de2014mental, van2015anonymity}. Meanwhile, on Yik Yak, design features such as ephemerality and hyper-locality have yielded different behavioral outcomes~\cite{Schlesinger2017SitutatedAnonymity}. 

Many tools sit at this digital democratic intersection and invite anonymous participation in decision making and government decisions. Examples of digital tools that have been designed to enable e-government and civic engagement anonymously include Pol.is~\cite{Polis}, CommunityPulse~\cite{CommunityPulse2021}, and CommunityCrit~\cite{Mahyar2018CommunityCrit}. Such tools have highlighted the value of community voice in informing civic decision making, and the unique ability of digital tools in gathering large amounts of community data, making sense of that data, and communicating patterns to inform decisions. Furthermore, in many high stakes context such as work environments, worker voices have been elevated to management to give feedback anonymously without retribution~\cite{Abdulgalimov2023employeevoice, estell2021affording}.

We contribute to this literature by exploring anonymization of the literal voice in audio of spoken stories. We look at listeners' perceptions of storytellers when their voices have been anonymized or not, as well as whether knowledge about whether the speech utterance has been anonymized affects their impressions. Finally, we evaluate not the behavioral outcome of people as anonymous actors, but rather the impact on their feelings of being heard by others when their voices have been anonymized.
\section{Application Setting: Technology-Enhanced Civic Dialogue Network}
\label{sec:civic_dialogue_network}

The work in this paper was done in the context of the Fora conversation platform developed by the nonprofit Cortico,\footnote{\url{https://cortico.ai/}} a \textit{technology-enhanced civic dialogue network} set in the United States. In this section, we briefly overview the structure and value of the civic dialogue network and the role of voice anonymization within it.

Throughout the United States and globally, there has been a decline in trust, increased polarization, and threats to democracy~\cite{fiorina2008political, pewresearchcenter_2023}. Some members of the public feel generally unheard, expressing dissatisfaction with the current political system as well as their agency and ability to meaningfully take part in broader social decisions that affect their lives~\cite{innes2004reframing}. The conditions of traditional means of civic participation such as voting and town halls often invite reductive, non-nuanced expressions of opinions that can restrict participation and perpetuate injustices~\cite{innes2004reframing}. Consequently, there have been many explorations into how to more effectively share peoples' voices and allow them to participate in civics with nuance and greater agency, in pursuit of a more just civic system and democracy. These explorations encompass a range of alternative forms of decision making, from citizens' assemblies~\cite{fournier2011citizens} to community organizing and movement building~\cite{ganz2009david, ganz2010leading, yang2016narrative} to technology-enhanced civic participation~\cite{bouzguenda2019towards, CommunityPulse2021, Mahyar2018CommunityCrit}.

Fora contributes to this field of explorations by attempting to increase feelings of agency and being heard within civic processes. It aims to do so via the following: 1) facilitated dialogue meant to increase understanding and connection between participants through nuanced story, question, and opinion sharing, 2) recording and systematically analyzing voices and themes from those dialogues to reveal patterns across and within communities around key issues, and 3) partnering with civic leaders committed to listening and acting upon these insights. 
The platform is run by a central non-profit organization which supports and trains civic leaders, community organizations, corporate spaces, schools, etc. to host facilitated dialogues within their communities around specific issues. Facilitated dialogue is an age-old method of gathering in which a figure with some authority, a facilitator, supports a gathered group through an experience-centered, structured conversation. Usually, there are norms that structure turn taking, types of contributions, the design of the space, and how to listen and engage with others. The facilitator often guides the conversation with the aid of a script or developed notes to elicit nuanced experiences, opinions, and questions from participants. These types of dialogues are a key method used in restorative justice spaces to heal harms, perform mediation across conflicts and differences, and deepen already existing relationships through Circle practice~\cite{morris2001restorative, ortega2016outcomes, baldwin2010circle}.

While the conversations themselves are of great value, a key component of Fora is that it enables content shared within the conversations to be heard elsewhere as well. By recording the conversations, analyzing them with participatory, qualitative methods, and sharing the emergent patterns through public-facing portals, specific highlighted stories and key themes expressed in the dialogues can be explored by a much larger public audience. Furthermore, many conversation campaigns are held in partnership with civic leadership who commit to listening, engaging, and responding to these voices. This means that dialogues can have a direct and visible path to practical impact and influence, potentially increasing participant feelings of being heard and having agency.

Central to the design of the network is the human voice. Participants verbally share their stories, which are often tied to life experiences; thus, any data point in the system is always tied to one or more audio recordings of participants' voices. In particular, a key aspect of the system is that, when a highlighted story is shared in a public portal, the audio of the participant's voice is always shared along with the text transcript (this is only done if the participant grants consent). This is an important and central design choice for the network because of the reasons outlined in Section \ref{related_work_1}. However, there are settings in which people may be reluctant to allow their voices to be made publicly available if they have shared sensitive content or fear retaliation against what they have said, even if their story is especially powerful or insightful. Furthermore, as outlined in Section \ref{related_work_2}, the general capability for anonymity can be a highly valuable tool in social and political spaces to allow people to participate fully and authentically. 

For these reasons, we explore how we might integrate voice anonymization into the Fora civic dialogue network. Specifically, we are interested in studying how to best maintain the rich information within the human voice, but transform it in a way that successfully hides the speaker's identity. Given that feelings of being heard and hearing others are key goals in the network, it is essential to evaluate how anonymization methods affect those goals or not. In the following sections, we outline our methods and results as we explore these questions through two studies.
\section{Methods}

We treat voice anonymization as the task of suppressing personally identifiable attributes of a speech signal while preserving the key attributes related to semantic content.
It is relatively simple to anonymize speech by modulating voice characteristics using signal processing techniques.
For example, one may alter the frequency spectrum to change the perceived pitch of the voice or design acoustic filters to change the spectral characteristics of the speech; many speech anonymization methods popularly used in the media today (e.g., for anonymous interviews) fall under these categories.
However, these methods often result in robotic, unnatural sounding transformed voices.
Because they are not explicitly designed to retain the comprehensibility of the speech, they can also necessitate the use of subtitles or captions to allow listeners to understand what is being said.
In this study, we were specifically interested in voice anonymization methods that did not have this drawback, as perceptual intelligibility and clarity for listeners are critical components for our civic dialogue network setting.
Therefore, we resorted to two speech transformation/synthesis technologies that satisfy our basic surface level requirement of preserving intelligibility: voice conversion (VC) and text-to-speech (TTS).

Voice conversion is the task of transforming a voice to sound like another person without changing the linguistic content of the original utterance~\cite{sisman2020overview}.
It belongs to the general field of speech synthesis, which includes TTS as well as the changing of other speech properties such as emotion and accents.
In our setting, we further define it as the task of changing a speech utterance's timbre (i.e., a speaker's vocal tone) while leaving all other aspects of the utterance such as prosody, rhythm, and accent unchanged~\cite{qian2019autovc}.
VC has been used as the key component of many previously proposed voice anonymization systems~\cite{pobar2014online, qian2017voicemask, fang2019speaker, han2020voice, patino2021speaker}.
Meanwhile, TTS is the task of producing synthetic speech that corresponds to the verbalization of a text input.
It is the core technology that drives the voices of many speech interfaces such as Amazon's Alexa, Apple's Siri, and Google Assistant.

One might naturally expect VC to better fulfill the desired characteristics of a voice anonymization system for our civic dialogue network, as it preserves both the linguistic and prosodic content of a speech utterance while only changing the perceived speaker identity.
Meanwhile, TTS only preserves linguistic content, removing prosodic characteristics and other paralinguistic information in addition to changing the speaker identity.
However, we sought to compare these two ``levels'' of anonymization and study the impact that their differences would bring.

To perform VC, we used LVC-VC XL~\cite{kang23b_interspeech}, a recently proposed  deep neural network-based voice conversion model.
The model was trained on the VCTK Corpus~\cite{yamagishi2019vctk}, which consists of around 44 hours of English speech from 109 speakers (62 female, 47 male) with various accents (American, Australian, Canadian, English, Indian, etc.).
The model is \textit{zero-shot}, which means that it is able to convert any \textit{source} speaker's voice to sound like any \textit{target} speaker's voice; this was a necessary characteristic because the model needed to be usable for any arbitrary speaker.
For simplicity, we limited our selection of candidate target speaker voices to the 109 speakers that the model had seen during training and randomly sampled speakers from the pool of candidates when performing anonymization.
To control for gender, we always selected target speakers that were of the same gender as the original speaker; further details on speaker selection are described in Sections \ref{subsec:study1} and \ref{subsec:study2}.
Note that LVC-VC only changes the source speaker's timbre---all other aspects of the source utterance, including rhythm, intonation, and accent, stay the same.

For TTS, we first performed automatic speech recognition (ASR) using Whisper-base~\cite{radford2023robust} to obtain the text transcript of the speech utterance.\footnote{The Whisper ASR model was downloaded and run locally on the authors' computing hardware, rather than using OpenAI's API.}
Then, we manually corrected any errors and formatted the transcript so as to match the rhythm and tempo of the original speech utterance as closely as possible (for example, adding a comma inserts a brief pause between words).
Finally, to synthesize speech, we used VITS~\cite{kim2021conditional}, a state-of-the-art multi-speaker neural network model, which was also trained on the VCTK Corpus.
Here, we limited our pool of voices to speakers with American accents, specifically General American English accents, which left us with 17 female voice candidates and 5 male voice candidates (22 total).
As with VC, we always performed anonymization by selecting speakers that were of the same gender as the original speaker.
Note that TTS candidate voices were not specifically selected to match the accents of the original speakers; while most of the original speech utterances in our studies were spoken in General American English accents (7/10 in Study 1 and 18/21 in Study 2), some were not.
It is possible that a perceived difference in speaker demographics could have affected listeners' perceptions, as has been observed and documented in prior studies~\cite{derwing2002teaching, russo2017non, fuse2018perception}.
However, it was unfortunately infeasible to completely control for these factors because of the limited set of accents that the TTS model could synthesize speech in (for example, it is not able to synthesize speech in Spanish or other Hispanic accents).
We leave addressing this as potential future work.
Further details on speaker selection for each study are described in Sections \ref{subsec:study1} and \ref{subsec:study2}.

We conducted two studies to measure the impact of voice anonymization when done using either VC or TTS.
In the first study ($n = 1500$), we investigated the impact of anonymization on \textit{listeners} of stories told in our civic dialogue network, measured in terms of the empathy and trust they felt towards the storyteller.
In the second study ($n = 21$), we investigated \textit{speakers'} perceptions of their own stories in the civic dialogue network that had been anonymized, as well how effectively they felt that their identity was masked.
Both study protocols were reviewed and approved by a university's institutional review board (IRB).
In the rest of this section, we describe the design, procedures, and analyses done for each of the two studies.

\subsection{Study 1: Perceptions of the Listener}
\label{subsec:study1}

In our first study, we investigated how anonymizing stories told in our civic dialogue network affected the perceptions of listeners towards the storyteller.
We also considered how knowledge of whether a story had been anonymized or not (i.e., telling a listener that the speech utterance had been altered for the purposes of anonymization) affected their responses.
Concretely, we aimed to answer the following research questions (RQs):

\begin{itemize}
    \item \textbf{RQ 1:} How does anonymizing speech impact a listener’s perception of and/or connection with the storyteller, especially in terms of trust and empathy?

    \item \textbf{RQ 2:} How does knowledge of whether speech has been altered for the purposes of anonymization affect a listener’s trust and empathy towards the storyteller?
\end{itemize}

\begin{table}
    \centering
    \caption{Topics, speaker genders, and durations of the original, VC, and TTS audio files for the 10 stories used in Study 1.}
    \resizebox{\linewidth}{!}{
        \begin{tabular}{clccc}
            \toprule
            \textbf{Story \#}    & \textbf{Topic} & \textbf{Gender}    & \textbf{\makecell{Duration \\ (Orig./VC)}}    & \textbf{\makecell{Duration \\ (TTS)}} \\
            \midrule
            1   & Rent, housing stability, and homelessness     &   Female  & 2:10  & 1:06 \\
            2   & Police brutality on a family member           &   Female  & 2:56  & 2:01 \\
            3   & Dental care for children                      &   Female  & 1:11  & 0:46 \\
            4   & Laws and support for trans rights             &   Female  & 1:36  & 1:00 \\
            5   & Life lessons learned from a mother            &   Female  & 1:17  & 0:47 \\
            6   & Parental pressure on children playing sports  &   Male    & 0:58  & 0:47 \\
            7   & Growing up, maturing, and building relationships with siblings    &   Male   & 1:39  & 1:06 \\
            8   & Neighborhood gun violence and its impact on youth                 &   Male   & 0:51  & 0:33 \\
            9   & Recognition of social privilege               &   Male    & 1:01  & 0:58 \\
            10  & Volunteering for people from a different social background        &   Male   & 2:10  & 1:06 \\
            \bottomrule
        \end{tabular}
    }
    \label{table:story_info}
\end{table}

\subsubsection{Procedure}

We collected audio clips of 10 stories from our civic dialogue network that had been spoken by a variety of American speakers across age, gender, and emotional valence.\footnote{The exact age breakdown of the speakers was unavailable because we took the stories from conversations where that information was not collected.}
The stories covered topics such as housing stability, police brutality, trans rights, sibling relationships, and neighborhood violence.
Five stories were spoken by women, while five were spoken by men.
The stories ranged from 51 seconds to 2 minutes and 56 seconds in length ($\mu = 89.6$ seconds, $\sigma = 36.3$ seconds).
More detailed information on each of the stories is shown in Table \ref{table:story_info}.

We generated anonymized versions of the stories by performing VC and TTS on the audio files as described above.
For VC, we randomly selected ten potential target speakers of the same gender as the storyteller from the VCTK Corpus and used them to generate voice converted candidates.
Then, we manually selected one VC candidate that sounded sufficiently different from the original speaker's voice while maintaining clear audio quality.
For TTS, we generated candidates using all of the American-accented voices from the VCTK Corpus that were of the same gender as the storyteller (17 for female, 5 for male) and manually selected the candidate that sounded the most natural.
Audio that had been anonymized using VC maintained the same length as the original speech utterances.
For TTS, the synthesized speech ranged from 33 seconds to 2 minutes and 1 second in length ($\mu = 61.2$ seconds, $\sigma = 22.6$ seconds); the shorter lengths reflected the lack of pauses, repetitions, and filler words that are part of regular human speaking patterns~\cite{brennan1995feeling} but were not included in the generated TTS audio.

The objective of this study was to measure participants' responses towards the VC and TTS voices and compare them against responses towards the original, un-anonymized voice.
Each study participant listened to a single random audio clip corresponding to one of three types of audio (Original, VC, or TTS) for one of the 10 selected stories.
Participants were also randomly assigned to an ``aware'' or ``unaware of anonymization'' condition; those in the ``aware'' conditions saw the following message before being shown the audio player:
\begin{quote}
    \textit{Please note that the voice you will be hearing has been altered in order to anonymize the identity of the speaker.}
\end{quote}
while those in the ``unaware'' condition were only shown the audio player with no other text beyond the instructions.
We included a check in the survey to ensure that all participants finished playing the audio before they could proceed.
Then, participants answered a series of 20 Likert scale questions that were designed to measure various aspects of their trust and empathy towards the storyteller in the audio clip, as well as the following free response question:
\begin{quote}
    \textit{What were some aspects of the audio clip that made you provide the answers that you did above?}
\end{quote}
Further details on the survey design are described in Section \ref{subsubsec:survey_design}.

Altogether, there were 6 audio type and knowledge conditions: 3 audio (Original, VC, and TTS), and 2 knowledge (aware, unaware).\footnote{Note that one combination of conditions, Original + aware, was a deceptive condition in which we told listeners that the audio had been altered although it was not.}
For each condition on each of the 10 stories, we collected 25 responses.
In total, this made for 1500 survey participants.

\subsubsection{Survey design}
\label{subsubsec:survey_design}

Given the key goals of our civic dialogue network, we looked to measure the empathy and trust of listeners towards storytellers after listening to audio of their stories.
To measure this, we created a 5 point Likert scale survey (1 -- strongly disagree, 5 -- strongly agree) measuring \textit{state empathy} and \textit{trust} of the listener towards the storyteller.
For empathy, we collected 12 survey questions from established research around state empathy~\cite{shen2010scale}.
We chose state empathy over other empathy measures as we were not interested in evaluating the general empathy of the listener, but rather their experience of empathy towards the speaker informed solely by the story they heard.
For trust, we developed a set of 8 questions jointly with a key conversation designer within the civic dialogue network.
We then validated with the conversation designer that both sets of questions on empathy and trust were measuring factors that were in alignment with the civic dialogue network's purposes and goals.
The survey questions are shown in Table \ref{tab:survey_1_questions}.

\begin{table}
    \centering
    \caption{Outline of the 20 Likert scale survey questions used in Study 1. Each set of questions was developed to address a different relevant theme, namely empathy or trust.}
    \resizebox{\linewidth}{!}{
        \begin{tabular}{ll}
            \toprule
            \textbf{Theme} &  \textbf{Question}\\
            \midrule
            \multirow{12}{6em}{Empathy} & \textbf{E1:} I believe the speaker’s emotions in the story were genuine.\\
            & \textbf{E2:} I experienced the same emotions as the speaker when listening to the story.\\
            & \textbf{E3:} I was in a similar emotional state as the speaker when listening to the story.\\
            & \textbf{E4:} I could feel the speaker’s emotions.\\
            & \textbf{E5:} I could see the speaker’s point of view.\\
            & \textbf{E6:} I recognized the speaker’s situation.\\
            & \textbf{E7:} I could understand what the speaker was going through in the story.\\
            & \textbf{E8:} The speaker’s reactions to the situation were understandable.\\
            & \textbf{E9:} When listening to the story, I was fully absorbed.\\
            & \textbf{E10:} I could relate to what the speaker was going through in the story.\\
            & \textbf{E11:} I could identify with the situation described in the story.\\
            & \textbf{E12:} I could identify with the characters in the story.\\
            \midrule
            \multirow{8}{6em}{Trust} & \textbf{T1:} I believe that the speaker was telling an authentic life experience.\\
            & \textbf{T2:} I believe the story is true.\\
            & \textbf{T3:} I felt like I could trust the speaker.\\
            & \textbf{T4:} I have respect for the speaker.\\
            & \textbf{T5:} I believe that the speaker wanted me to understand their experience and perspective.\\
            & \textbf{T6:} I believe the speaker was manipulating me or trying to persuade me with their story.\\
            & \textbf{T7:} I would like to learn more about the speaker.\\
            & \textbf{T8:} I would be willing to talk with the speaker and share some of my own experiences.\\
            \bottomrule
        \end{tabular}
    }
    \label{tab:survey_1_questions}
\end{table}

\subsubsection{Participants}

We recruited all participants on the Prolific crowdsourcing platform.
Participants were required to be located in the United States or United Kingdom and be fluent in English.
53.33\% of participants identified as female, 46.40\% identified as male, and the remaining 0.27\% preferred not to say.
The racial breakdown of participants was 84.93\% White, 5.8\% Asian, 4.07\% Black, 3.8\% Mixed, and 1.4\% Other.
Participants were paid \$1 USD for their time.
The median time taken for the task was 4 minutes and 6 seconds, which resulted in a reward per hour of \$14.63 USD.

\subsection{Study 2: Perceptions of the Speaker}
\label{subsec:study2}

In our second study, we looked to measure speakers’ perceptions of their own speech that had been anonymized via VC or TTS. The research questions were:

\begin{itemize}
    \item \textbf{RQ 1:} How does anonymizing speech affect a speaker’s own impression of how their emotions and intended message are conveyed?
    
    \item \textbf{RQ 2:} How do different speech anonymization methods affect a speaker’s perception of their privacy?
\end{itemize}

\subsubsection{Participants and procedure}

To understand the impact of voice anonymization on speakers' perceptions of their own stories and the degree to which they felt heard, we reached out to a pool of participants from one set of conversations that had been held using the civic dialogue network described in Section \ref{sec:civic_dialogue_network}.
Given that the application of our research deals with the challenges of feeling agency and being heard within civic spaces, it was important that we looked to participants of real-world conversations that had actual consequences; without the weight of dealing with real stories, people may have responded superficially to our study.
Furthermore, as a key aspect of voice anonymization is the protection of privacy, it was important that we choose participants and stories for which privacy was relevant to some degree.
Therefore, we worked with a set of conversations that was held within a university's work environment, in which participants discussed both the joys and the frictions that were part of their workplace.

We reached out to all participants who took part in at least one of two such conversation campaigns within their workplace, which made up a pool of 82 people.
A total of 21 participants completed our survey (see Section \ref{subsubsec:study2_survey_design}).
71.43\% of participants identified as female, 23.81\% identified as male, and 4.76\% identified as non-binary.
The racial breakdown of participants was 61.9\% White, 19.05\% Hispanic or Latino, and 9.52\% Asian.
Participants were compensated for their involvement in the study by being entered into a lottery with the chance to win a \$50, \$35, or \$20 gift card.

\subsubsection{Survey design}
\label{subsubsec:study2_survey_design}

The survey for this study aimed to measure how storytellers perceived their own stories and voices that had been anonymized using VC and TTS.
Specifically, we evaluated how anonymizing the audio of stories using the two methods affected storytellers' perceptions of conveyed emotions, their feelings of being heard, and their feelings of security and privacy when sharing their stories with others.

To do this, we first had participants listen to their own story in the original, unaltered voice.
Then, they were asked to imagine a hypothetical scenario in which their stories would be shared through a public-facing digital portal that would be visible to all members of their broader workplace community, including their peers and superiors.
Under this condition, they were told to consider the situation of sharing an anonymized version of their story on the portal.
Participants then listened to both VC and TTS versions of their story that we generated using the following procedure.
For VC, we generated four candidates per story using 2 female and 2 male voices randomly chosen from the VCTK Corpus as target speakers; we presented all four options to the participant and asked them to choose which version they would most like to use in the hypothetical scenario.
For TTS, we followed the same procedure as in Study 1, where we generated candidates using the voices in American accents from the VCTK Corpus that were of the same gender as the participant.
Then, we manually selected the single option whose speech sounded the most natural.

After listening to the VC and TTS candidates, participants answered the following 5 point Likert scale questions (1 -- strongly disagree, 5 -- strongly agree) in response to each type of audio:
\textit{
\begin{enumerate}
    \item I feel like my story would still be heard with the \textnormal{(transformed, synthesized)} voice.
    \item I feel like the \textnormal{(transformed, synthesized)} voice retained the authenticity of my story.
    \item I feel like the \textnormal{(transformed, synthesized)} voice conveys the emotion I wanted to convey. 
    \item I feel like the \textnormal{(transformed, synthesized)} voice retains important characteristics of my voice (accent, speech pattern, intonation, etc.).
    \item I feel like the \textnormal{(transformed, synthesized)} voice masked my identity.
\end{enumerate}
}
as well as the following free response question:
\begin{quote}
    \textit{Please provide any additional comments on your impressions of the \textnormal{(voice conversion, text-to-speech)} technology applied on your voice in the context of \textnormal{[your community]}.}
\end{quote}

This was then followed by a question asking the participant about the members of their community with whom they would be comfortable sharing their story in the original, VC, and TTS voices.
The question was posed as a checkbox selection question where multiple selections were possible, with the following options:
\begin{itemize}
    \item \textbf{Group 1:} My community outside of my workplace (friends, acquaintances, family)
    \item \textbf{Group 2:} My peers (other graduate students, other staff members, whatever community you are a part of)
    \item \textbf{Group 3:} The broader university community
    \item \textbf{Group 4:} My supervisor(s)
    \item \textbf{Group 5:} University administration
    \item \textbf{Group 6:} People who have power over me
    \item \textbf{None of the above}
\end{itemize}
The survey finished with a question asking if their answers changed across the three conditions (Original, VC, and TTS), and if so, to elaborate on why.

\subsection{Data Analysis Methods}
\label{subsec:data_analysis}

\subsubsection{Likert scale questions}

We performed two-sided Student's $t$-tests to compare the responses to the Likert scale questions between the various conditions.
For Study 1, we performed $t$-tests to compare responses for Original vs. VC and Original vs. TTS for all 20 questions.
We also performed tests to compare the responses for each audio type (Original, VC, TTS) when listeners were aware vs. unaware of anonymization.
For Study 2, we performed $t$-tests comparing the speakers' responses for the VC and TTS audio.

\subsubsection{Free response questions}
\label{subsubsec:frq_analysis}

Because of the large number of responses to the free response question in Study 1, we performed a topic analysis of the answers rather than manually going through and analyzing them.
To do this, we leveraged Sentence-BERT~\cite{reimers2019sentence}, a neural network language model that has been trained to encode sentences into semantically meaningful representations.
Sentence-BERT maps sequences of words (usually a sentence or phrase) to 768-dimensional vector representations (henceforth referred to as “embeddings”).
Intuitively, the model maps semantically similar sentences or phrases to embeddings that are close to each other and semantically different ones to embeddings that are farther from each other in the latent vector space.

We split all 1500 free response answers into sentences using the Python Natural Language Toolkit (NLTK) package, producing 2024 sentences.
The full responses had an average length of 108 characters, and individual sentences had an average length of 80 characters.
Then, we computed embeddings for the sentences and clustered them using the Python implementation of HDBSCAN\footnote{\url{https://hdbscan.readthedocs.io/en/latest/}}~\cite{mcinnes2017hdbscan} with a minimum cluster size of 2.
Setting the minimum cluster size to 2 ensured that any response that was not in a cluster on its own would be represented as a topic/theme.
This resulted in 109 initial clusters formed by 928 responses (HDBSCAN also produced a null cluster containing 1096 elements).
Then, we went through the clusters and agglomeratively merged similar ones until there remained 20 groups of responses that represented a distinct set of emergent topics and themes.
We describe the results of this analysis in Section \ref{subsubsec:study1_subjective}.

\section{Results}
\label{sec:results}

In this section, we summarize our findings from the two studies.
Section \ref{subsec:results_study1} covers the results from Study 1, while Section \ref{subsec:results_study2} covers the results from Study 2.
In each subsection, we first analyze quantitative results taken from the Likert scale questions, followed by a more qualitative analysis of the answers to the free response questions.

\begin{figure}[t]
    \centering
    \begin{subfigure}[b]{1.0\textwidth}
        \centering
        \includegraphics[width=0.98\textwidth]{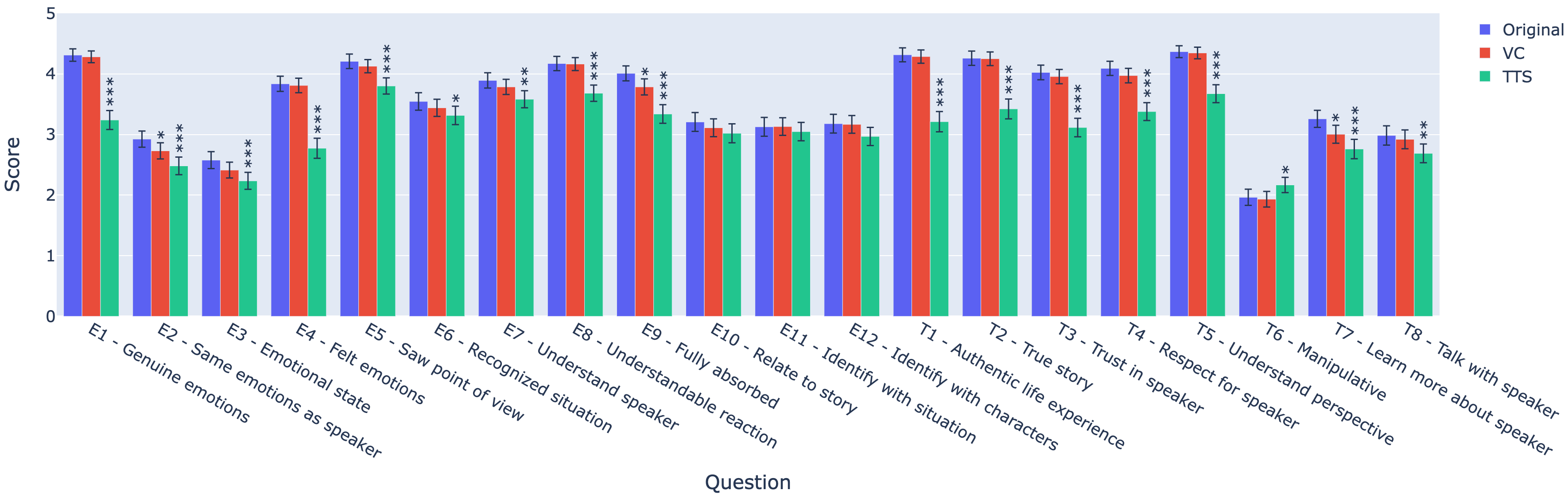}
        \caption{Unaware}
        \label{subfig:orig_vc_tts_unaware}
    \end{subfigure}
    \begin{subfigure}[b]{1.0\textwidth}
        \centering
        \includegraphics[width=0.98\textwidth]{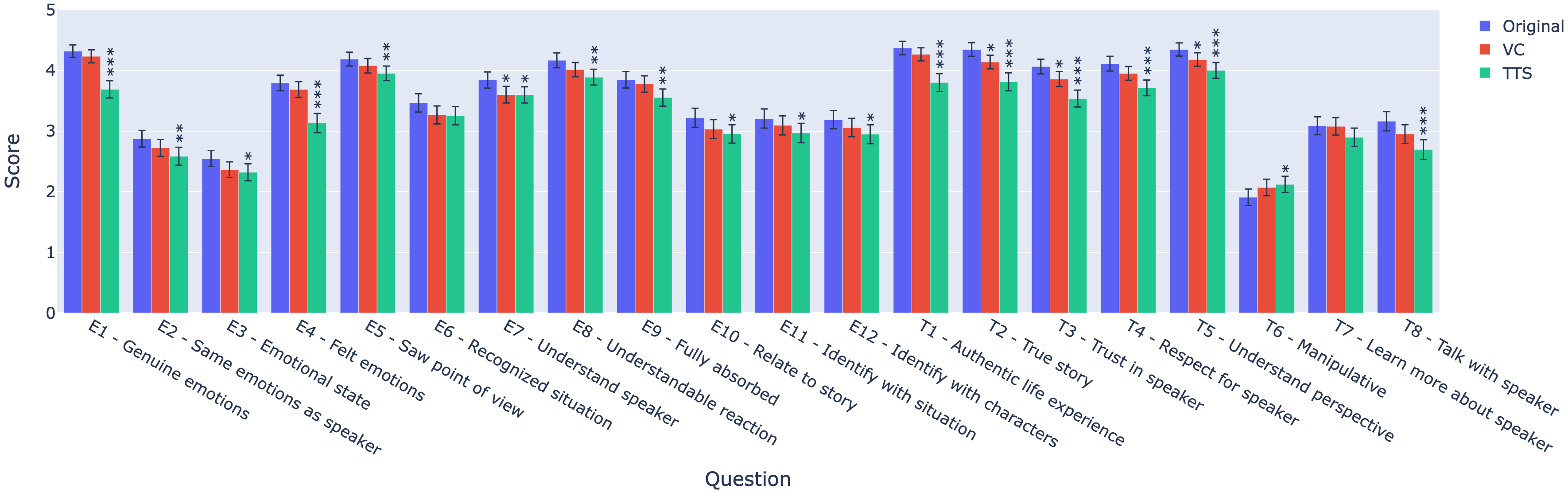}
        \caption{Aware}
        \label{subfig:orig_vc_tts_aware}
    \end{subfigure}
    \caption{Average scores for the 20 Likert scale questions in the survey for Study 1 when listeners were (a) unaware and (b) aware of anonymization. Error bars represent 95\% confidence intervals. *, **, or *** on top of bars for VC and TTS denote statistically significant differences at $p < 0.05$, $p < 0.01$, or $p < 0.001$, respectively, for two-sided $t$-tests comparing against scores for Original.}
    \label{fig:study1_scores}
\end{figure}

\begin{figure}
    \centering
    \begin{subfigure}[b]{1.0\textwidth}
        \centering
        \includegraphics[width=0.9\textwidth]{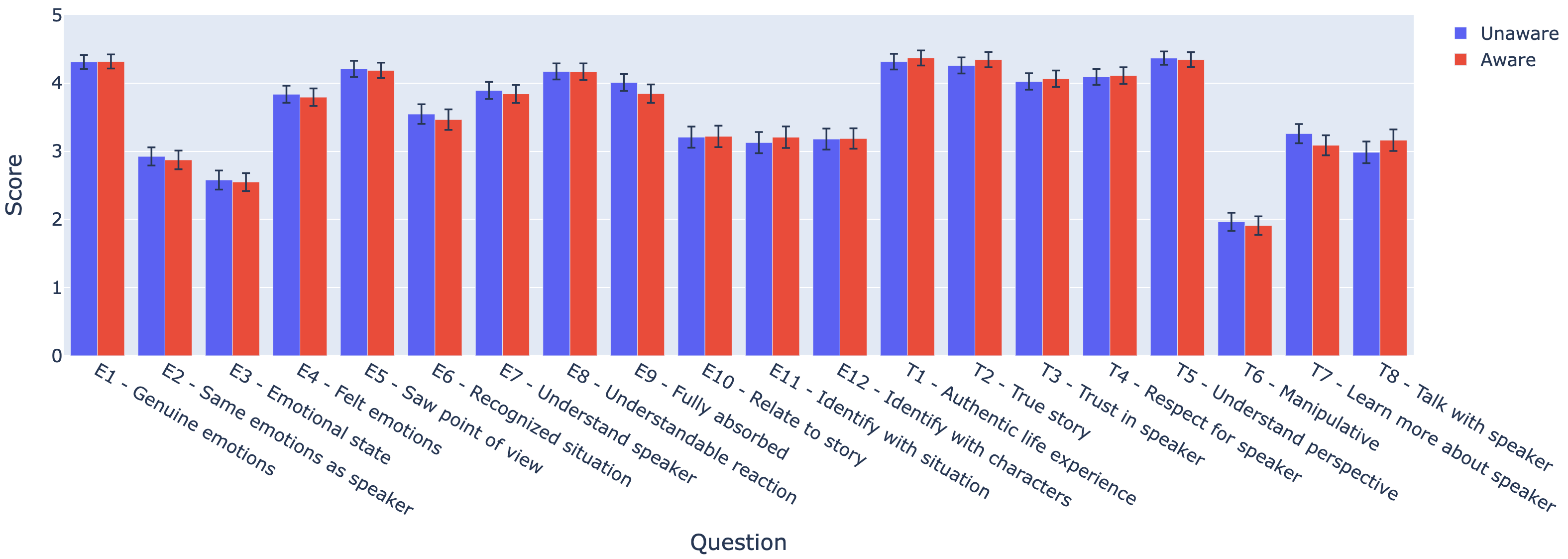}
        \caption{Original}
        \label{subfig:orig_awareness}
    \end{subfigure}
    \begin{subfigure}[b]{1.0\textwidth}
        \centering
        \includegraphics[width=0.9\textwidth]{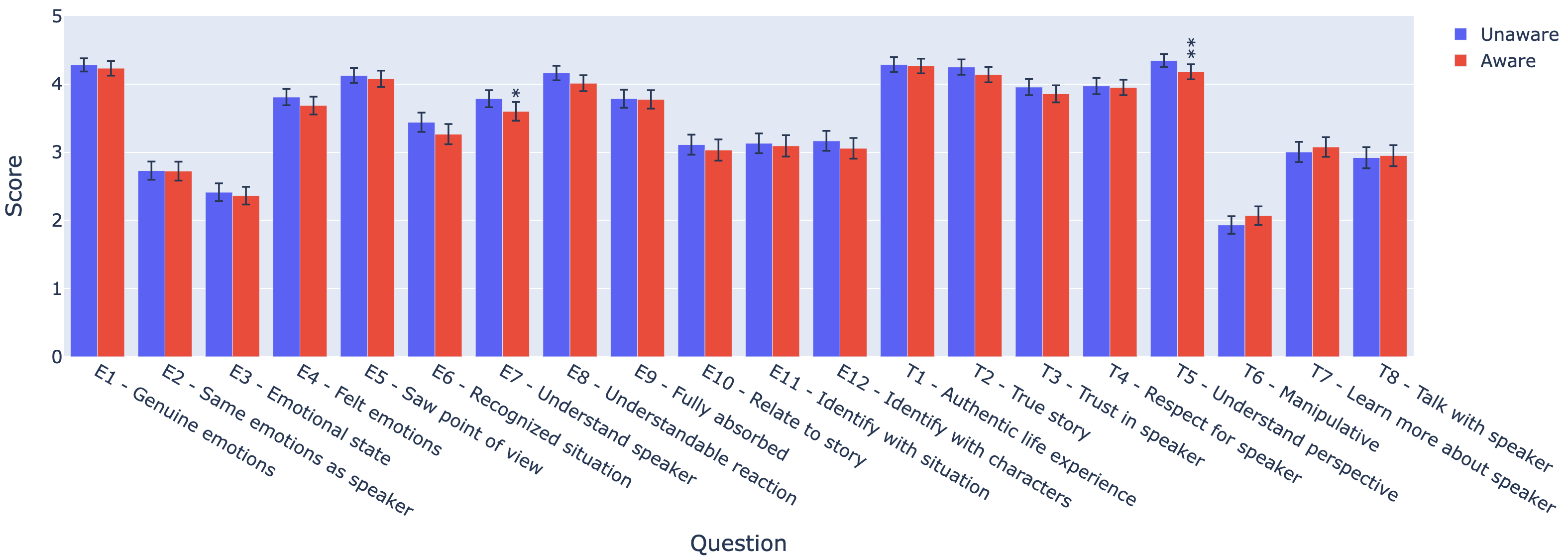}
        \caption{VC}
        \label{subfig:vc_awareness}
    \end{subfigure}
    \begin{subfigure}[b]{1.0\textwidth}
        \centering
        \includegraphics[width=0.9\textwidth]{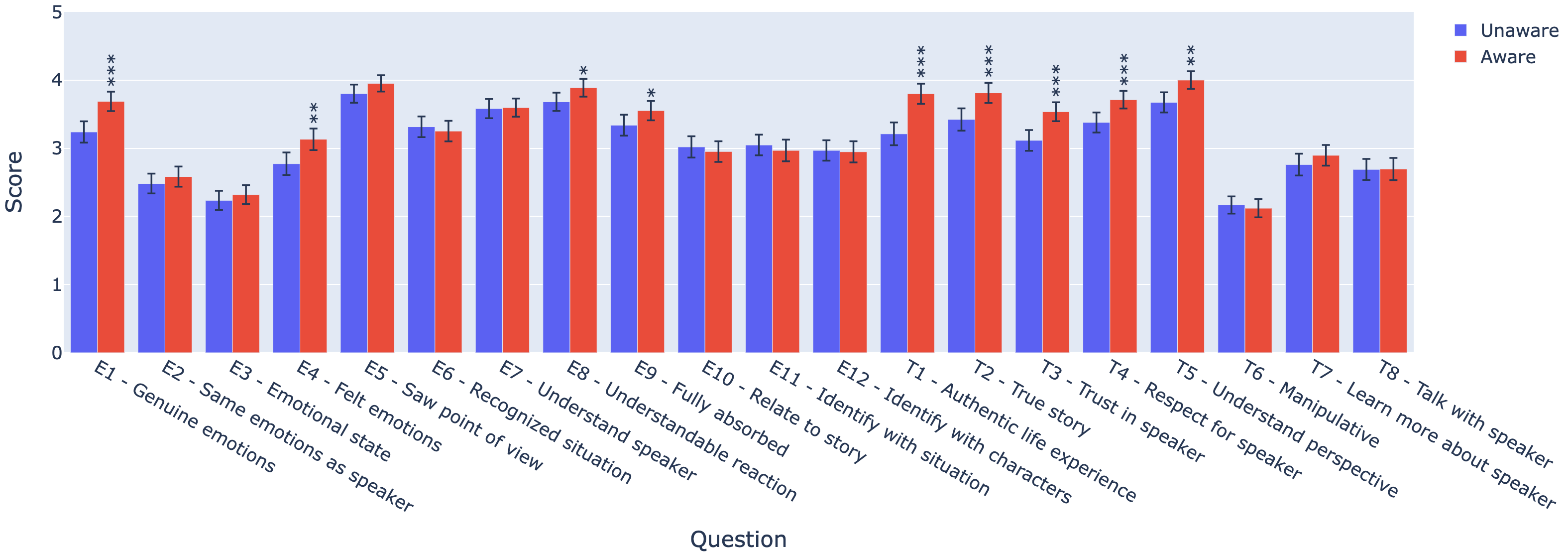}
        \caption{TTS}
        \label{subfig:tts_awareness}
    \end{subfigure}
    \caption{Average scores for the 20 Likert scale questions in the survey for Study 1 comparing the unaware (blue) and aware (red) conditions for (a) Original, (b) VC, and (c) TTS voices. Error bars represent 95\% confidence intervals. *, **, or *** on top of bars for Aware denote statistically significant differences at $p < 0.05$, $p < 0.01$, or $p < 0.001$, respectively, for two-sided $t$-tests comparing against scores for Unaware.}
    \label{fig:study1_awareness}
\end{figure}

\subsection{Study 1}
\label{subsec:results_study1}

\subsubsection{Listeners' perspectives on voice conversion vs. TTS for anonymization}
\label{subsubsec:results_study1_1}

We first analyze the perceptions that listeners had towards stories that were anonymized using VC or TTS when compared against a baseline of the original, unaltered voice.
Fig. \ref{fig:study1_scores} shows bar charts of the average scores for each of the 20 Likert scale questions in the survey, along with 95\% confidence intervals, when listeners were (a) unaware and (b) aware of anonymization.
Statistically significant differences between scores for the Original vs. VC/TTS conditions are denoted by asterisks above the bars.
For figure compactness, we label questions on the $x$-axis with abbreviated codes (E1, E2, etc.); the mapping back to the full text of the question can be found in Table \ref{tab:survey_1_questions}.

We find that stories anonymized using VC largely achieve similar scores to Original audio under both unaware and aware conditions.
When listeners were unaware of anonymization, there were statistically significant differences ($p < 0.05$) between Original and VC audio for only 3/20 questions; when listeners were aware of anonymization, there were statistically significant differences for only 4/20 questions.
This indicates that VC is successful at meeting our objectives for anonymization within the civic dialogue network framework; that is, stories are conveyed in the same way as if listeners hear the original voice, and there is little to no difference in terms of the empathy or trust that listeners feel towards the storyteller.

Meanwhile, stories anonymized using TTS show significantly different scores than Original audio; there are statistically significant differences in 17/20 and 18/20 questions under the unaware and aware conditions, respectively.
Notably, stories spoken using the TTS voice result in significantly lower scores in terms of perceived emotion (E1, E4), as well as trust, respect, and perception of authenticity (T1, T2, T3, T4, T5).

\subsubsection{Awareness of anonymization}

We now analyze the impact that awareness of anonymization has on listeners' empathy and trust for a storyteller.
Fig. \ref{fig:study1_awareness} shows scores for the Likert scale questions when performing head-to-head comparisons between the unaware and aware conditions for (a) Original, (b) VC, and (c) TTS voices.
We again find largely similar patterns between the Original and VC voices.
For the Original voice, there are no statistically significant differences between the two conditions for any of the 20 questions, while for the VC voice, there are statistically significant differences in only 2/20 questions.
These results further support the notion that VC voices largely share the same characteristics as original, unaltered voices when perceived by listeners.

When looking at the scores for the TTS voice, however, we see an interesting phenomenon.
When listeners are told that a TTS voice has been altered for the purposes of anonymization, they score the voice significantly higher in terms of perceived emotion (E1, E4), as well as trust, respect, and perceived authenticity (T1, T2, T3, T4, T5).
Note that these are precisely the categories and questions in which TTS lagged behind the Original and VC voices in Section \ref{subsubsec:results_study1_1}.
This suggests that when people know that a voice has been altered, they become less sensitive to the vocal qualities of the storyteller, perhaps because they may guess that the voice was artifically synthesized (as was the case here).
Consequently, the content of the story, rather than the delivery, may play a larger role in determining the listener's empathy and trust towards the storyteller.

\subsubsection{Free responses}
\label{subsubsec:study1_subjective}

The analysis of the answers to the free response questions described in Section \ref{subsubsec:frq_analysis} yielded a set of 20 key topics and themes, which are listed in Table \ref{tab:survey_1_frq_themes} along with the number of sentences from the responses belonging to each.
These provide insights into the factors that participants in our survey were considering when they provided their answers to the Likert scale questions.

Participants generally cited positive sentiments when they provided high scores for empathy and trust, such as ``genuine, real sounding experience'' or ``perceived passion and sincerity.''
One participant wrote, ``I felt I could hear how passionate she was about her community, and heard her voice break when she spoke about her frustrations and disappointment that the state she lives in had found a new scapegoat.''
Another key factor seemed to be sympathy for the speaker that was elicited by their voice: ``I could just feel the emotions in her voice, and although I may not have ever had any similar experiences in my life, I could feel considerable empathy for her.''
Notably, some participants mentioned that relatability of the story was a key factor for empathy, but trust could be determined more so by the voice and speaking style: ``It wasn't a story that I could particularly relate to, but it seemed genuine and I had no reason to believe the speaker was inventing the story.''
Overall, these responses were most often connected to stories told using the Original or VC voices.

Meanwhile, lower empathy and trust scores were often linked to more neutral or negative sentiments, such as that the ``voice sounded not genuine or unconvincing'' or that the story ``sounded scripted'' or ``like an AI.''
One participant responding to a TTS+unaware story stated, ``judging from their voice, there wasn't much emotion there and I believe it was a mostly made up story,'' indicating that some listeners liken unnaturally un-emotive voices with insincerity or lower trust, regardless of the actual content of the story.
However, other speakers were able to look past a lack of emotion and focus on the story's content: ``I felt like the person's voice was pretty monotone and didn't convey emotion, but I do understand the situation the speaker is in and could identify with and empathize with that.''
Interestingly, the TTS+aware condition yielded some responses like: ``I could hear the emotion in her voice as she told her story, she sounded very genuine,'' and ``The tone of voice and emotion when telling the story was authentic.''
These are sentiments that were rarely expressed in the TTS+unaware condition, and the responses possibly indicate that awareness of anonymization may have caused some listeners to subconsciously recalibrate their standards for expressivity and emotion.

\begin{table}
    \centering
    \caption{Topics and themes that emerged from answers to the free response question in Study 1 that asked, ``What were some aspects of the audio clip that made you provide the answers that you did above?'' The number of sentences from the free response answers belonging to each category are shown in parentheses.}
    \resizebox{\linewidth}{!}{
        \begin{tabular}{ll}
            \toprule
            \multicolumn{2}{c}{\textbf{Topics and Themes}}\\
            \midrule
            1. Genuine, real sounding experience       \textcolor{black}{(111)}  & 11. Relatable issue/content                           \textcolor{black}{(83)} \\
            2. Speaker sounds realistic and believable \textcolor{black}{(16)}   & 12. Unrelatable issue/content                         \textcolor{black}{(19)} \\
            3. Strong emotion in the voice             \textcolor{black}{(106)}  & 13. Did not know enough about the speaker             \textcolor{black}{(4)}  \\
            4. Tone of voice                           \textcolor{black}{(81)}   & 14. Voice sounded not genuine or unconvincing         \textcolor{black}{(20)} \\
            5. Perceived passion and sincerity         \textcolor{black}{(46)}   & 15. Monotone voice, lack of emotion                   \textcolor{black}{(94)} \\
            6. Speaker spoke clearly and naturally     \textcolor{black}{(22)}   & 16. Skeptical that the experience happened            \textcolor{black}{(13)}  \\
            7. Powerful life experience                \textcolor{black}{(9)}    & 17. Story unclear or speaker difficult to understand  \textcolor{black}{(88)} \\
            8. Sympathy for the speaker                \textcolor{black}{(18)}   & 18. Unclear audio                                     \textcolor{black}{(8)} \\
            9. Sympathetic but unrelatable story       \textcolor{black}{(9)}    & 19. Sounds scripted / like reading from a script      \textcolor{black}{(30)} \\
            10. Overall content of story               \textcolor{black}{(109)}  & 20. Doesn't sound like a real person / sounds like AI \textcolor{black}{(42)} \\
            \bottomrule
        \end{tabular}
    }
    \label{tab:survey_1_frq_themes}
\end{table}

\subsection{Study 2}
\label{subsec:results_study2}

\subsubsection{Storytellers' perspectives on voice conversion vs. TTS for anonymization}

\begin{figure}
    \centering
    \includegraphics[width=0.8\textwidth]{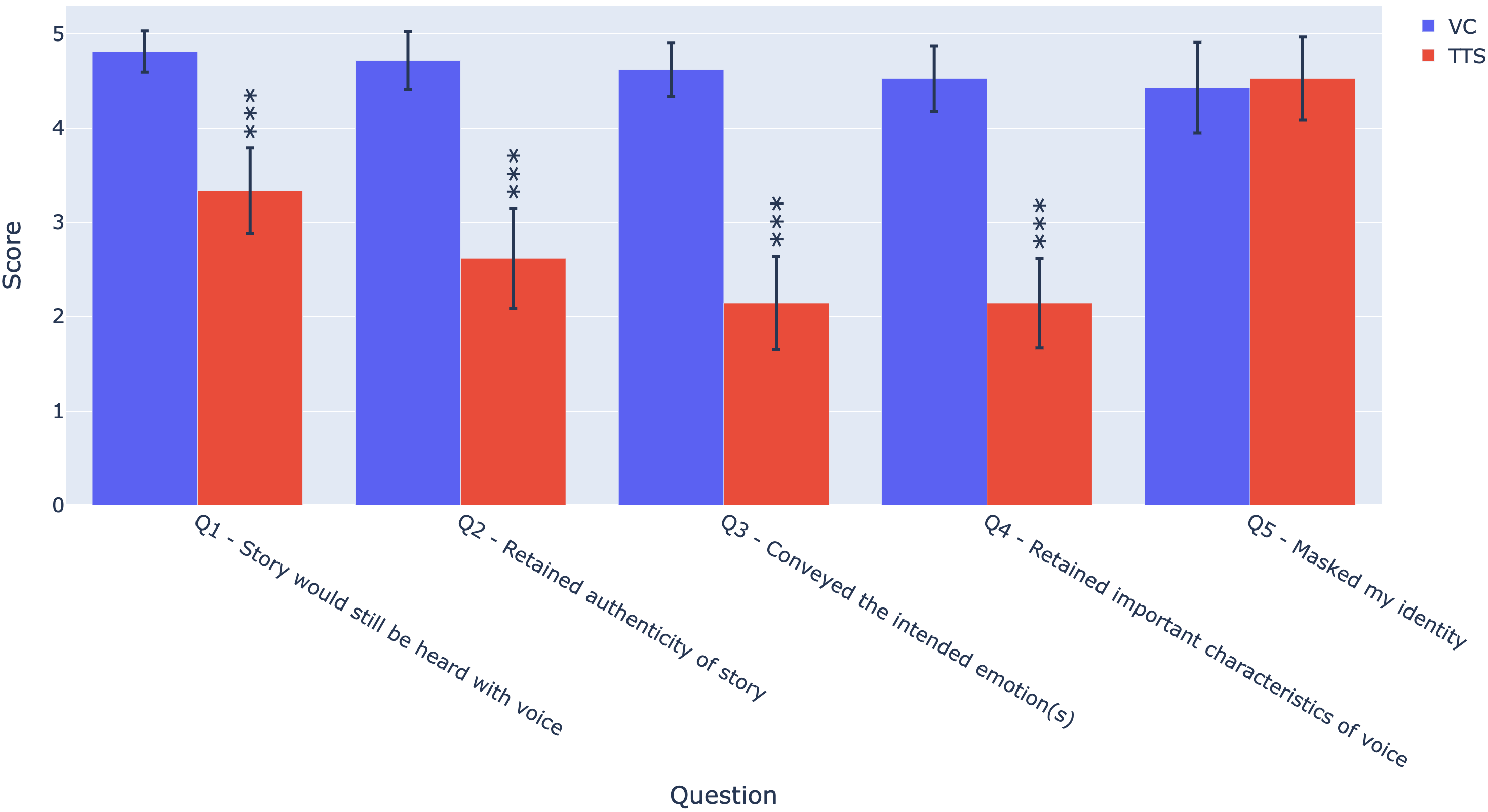}
    \caption{Average scores for the 5 Likert scale questions in the survey for Study 2 comparing the perceptions of speakers towards their own voices that were anonymized using VC (blue) and TTS (red). Error bars represent 95\% confidence intervals. *, **, or *** on top of bars for TTS denote statistically significant differences at $p < 0.05$, $p < 0.01$, or $p < 0.001$, respectively, for two-sided $t$-tests comparing against scores for VC.}
    \label{fig:study2}
\end{figure}

Fig. \ref{fig:study2} shows a bar chart of average scores for the 5 Likert scale questions in the survey for Study 2, along with 95\% confidence intervals.
A clear pattern emerges when we analyze the perceptions of original storytellers towards stories that were anonymized using VC or TTS: VC voices consistently make speakers felt more heard, that the authenticity of their story would be retained, that the voice conveys the intended emotions, and that the anonymized voice retains important characteristics of their own voice.
This is perhaps unsurprising given the additional prosodic information that VC preserves compared to TTS.
However, the extent to which speakers show their preference is quite notable, as are the high scores that VC achieves.
These results show that VC can successfully maintain the rich paralinguistic qualities of speech from the perspective of original speakers as well, fulfilling one of the key requirements for our civic dialogue network setting.

\subsubsection{Speakers' perceptions of privacy}

Participants rated both the VC and TTS voices highly for their perception of how successful they were at masking their identities; indeed, it was the only Likert scale question in Study 2 for which the two conditions did not have a statistically significant different score.
Accordingly, when we looked at the community members with whom participants said they would be comfortable sharing their story with, we found that all participants marked exactly the same groups for versions of their story anonymized using both VC and TTS.
Participants responded that they would be comfortable sharing their story in a VC or TTS voice with an average of $\mu = 4.95$ groups ($\sigma = 1.69$).
This was a marked increase compared to the average number of groups that participants were comfortable sharing their story with in their original voice, which was $\mu = 2.95$ ($\sigma = 1.71$).
Notably, 13 out of 21 participants stated that they would be comfortable sharing their story with all six groups mentioned in the survey (excluding ``None of the above''), with a further two more (15 out of 21) stating that they would be comfortable sharing with every group except for one.
This was in contrast to the 3 out of 21 participants who were comfortable sharing their story with all six groups in their original voice.

\section{Discussion}

We have presented our findings from two studies: one measuring the impact of voice anonymization on listeners' feelings of empathy and trust towards individuals sharing personal stories, and one measuring the impact of voice anonymization on storytellers' own perceptions of being heard.
Through these studies, we have shown that voice conversion is a technology that can be suitable for real settings that require anonymizing speech: it produces speech that has almost no difference with real voices in terms of listener perception, it preserves the emotion and prosody-related characteristics from the original speech, and most importantly, it can successfully mask a speaker's identity given an appropriate target speaker to convert the voice to.
We have also shown that speech generated using TTS synthesis lags behind real and voice converted speech in these regards.
However, our observation that awareness of speech modification significantly changes the way in which listeners perceive TTS voices opens up several interesting questions for how we might design speech interfaces that utilize this technology.
For example, how might priming users of speech interfaces in different ways change how they react to the emotion and prosody, or lack thereof, expressed in TTS voices?

We additionally believe that our findings have meaningful implications for the design of technology-enhanced methods for civic dialogue.
While voice is often used only as a metaphor for civic participation, there are cases in which spoken language can be a significant means of participation in civic systems, for example, as outlined in the civic dialogue network we considered in this work.
Yet, mediums for richer communication often sacrifice opportunities for anonymity in exchange for the richness of the discourse~\cite{daft1986organizational}.
In scenarios where spoken language is the primary communication medium, we show that anonymizing speech using voice conversion can be a legitimate and desirable way of allowing individuals to preserve richness and nuance in stories without sacrificing their ability to participate anonymously.
Indeed, in numerous real-world applications of the civic dialogue network, we have repeatedly observed requests for voice anonymization techniques that can fulfill this purpose, coming from communities ranging from company workplaces to academic institutions including high schools and universities.
The widespread demand for these types of anonymization systems, whether in civic systems or beyond, can be further highlighted by the recent surge of work in this area, which has been spurred by initiatives such as the VoicePrivacy Challenge~\cite{tomashenko2022voiceprivacy, tomashenko2024voiceprivacy}.

Finally, we contribute to growing research within science, technology, and society studies as we explore not the effect of anonymity on behavior~\cite{boyd2012politics, donath2002identity, turkle2011life, de2014mental}, but its effect on trust and empathy from listeners and storytellers' perceptions of being heard.
These characteristics are unique and important in public discourse, especially in today's civic era in which we see an increase in distrust and weak feelings of agency~\cite{innes2004reframing}.
Given that anonymity can create conditions for discourse and deliberation free from coercion, our work takes a small step towards achieving the idealized vision of a liberal public sphere and a democracy filled with more equal, meaningful public discourse~\cite{habermas1991structural}. 

Computer-supported means of civic participation point to the desire for civic leaders and participants to have more nuanced, full opinions within their contributions~\cite{CommunityPulse2021, Mahyar2018CommunityCrit}.
However, most explorations into anonymous participation are often held back by limited, survey-style evaluation standards.
While we see some technologies such as Pol.is~\cite{Polis} experiment with anonymous discourse through chat, few look at anonymous discourse through voice, which further adds depth, richness, and nuance to one's civic contributions~\cite{daft1986organizational, schroeder2017humanizing}.
We hope that our work may spark a discussion around more exploratory ways for engaging in civic technologies that invite vocal participation in pursuit of improving our public sphere.
\section{Limitations}

We finish by outlining some of the limitations of the present work.
There are naturally scenarios in which neither voice conversion nor TTS are appropriate for performing anonymization.
For example, certain voices in a social group may still be identifiable by their accent or idiolect if anonymization is done using voice conversion, since the technology as presented in this work entirely preserves a speaker's accent and speaking style.
TTS offers an option for a different level of anonymization that could address some of these issues, and while we have seen that it removes the emotion and prosody-related characteristics from the original speech, it may still have a use case in situations in which voice conversion does not suffice.
Leveraging other recent developments in speech synthesis technologies, such as accent conversion~\cite{jin2023voice} or emotional TTS~\cite{um2020emotional}, may also help in this regard.
On the other hand, changing speech characteristics such as accent or idiolect may affect listeners' perceptions of a story entirely.
Any voice anonymization system should be chosen and implemented with care, considering the requirements and needs of the speakers, listeners, and stakeholders in the dialogue setting.

Social desirability bias, particularly in sensitive areas such as assessing one's warmth or empathy toward personal narratives, poses a challenge to the scientific integrity of studies like ours. To address this concern, we implemented several strategies to mitigate potential biases. First, we ensured randomization and uniformity across all survey conditions, thereby distributing any bias evenly across different stories. Additionally, participant responses were anonymized to promote candid feedback, and certain Likert-scale questions were reverse-coded to counteract response tendencies. However, it is important to acknowledge that our findings may still be influenced by an inherent bias toward higher empathy scores. Future research may address alternative methods for gauging empathy, such as eliciting responses based on how others might feel rather than solely focusing on participants' self-reported emotions.

In Study 2, conversations within a university setting were evaluated, involving faculty, staff, administrators, and students, encompassing diverse power dynamics and social complexities. However, storytellers represented only a narrow segment of the broader, complex U.S. population, potentially limiting the study's generalizability. While Study 1 participants offered broader representation, their stories remained confined within a U.S. context as well, again potentially raising questions of generalizability. Future research should explore applicability across diverse cultural contexts and representative samples beyond the U.S. context.
\section{Conclusion}

In this work, we explored how to effectively incorporate a voice anonymization system into a spoken language-based civic dialogue network whose purpose is to increase participants' feelings of agency and being heard within civic processes.
We tested two different speech transformation/synthesis methods for anonymization: voice conversion (VC) and text-to-speech (TTS).
In contrast to previous work, we explored the effect of anonymity not on user behavior, but on empathy and trust elicited from listeners of stories, as well as on speakers' own perceptions of being heard.
We found that voice conversion is a particularly suitable method for performing anonymization in our framework, as voice converted speech demonstrates almost no differences with real speech in terms of listeners' perceptions while successfully preserving the prosodic and other paralinguistic characteristics that were intended to be conveyed by the speaker.
Our work contributes to literature on applications of speech-based technologies as well as on exploratory methods for engaging in discourse, and we hope that it may be a small step towards greater civic participation and improvement of our public sphere.

\bibliographystyle{ACM-Reference-Format}
\bibliography{references}

\end{document}